\documentstyle[12pt,epsf]{article}
\newcommand{\beq}{\begin{equation}}
\newcommand{\eeq}{\end{equation}}
\newcommand{\ppd}{\partial}

\begin{document}

\begin{center}
{\bf Soliton solutions of Chiral Born-Infeld Theory and baryons.}
\end{center}

\begin{center}
O.V. Pavlovski\u{\i}
\end{center}

\begin{center}
{\em Institute for Theoretical 
Problems of Microphysics,}  {\em Moscow State University 
} \\ {\em Moscow, 119992, Russian Federation.}
\end{center}

$$
$$

\begin{center}
{\bf Abstract.}
\end{center}
$$
$$
{\small 
Finite-energy topological spherically symmetrical solutions of Chiral 
Born-Infeld Theory are studied. Properties of these solution are obtained, 
and a possible physical interpretation is also given. We compute static 
properties of baryons (mass,main radius, magnetic main radius, 
axial coupling constant) whose solutions can be interpreted as the 
baryons of QCD.}

$$
$$

The concept of the baryon as a chiral  soliton  has a 
long history. The idea of  a unified theory for baryons and mesons that 
is formulated in terms of chiral field  was only proposed for 
the first time by Skyrme
\cite{sk}.  
This theory has a set of topological static solutions and one can classify 
these solutions by the value of the topological (baryon) charge. 
In this 
model such solutions are associated with baryon states with different baryon 
charges, and soliton with $B=1$ (skyrmeon) is treated as a nucleon. Such 
kind of solutions and quantum fluctuations about them is well studied now 
and results of such analysis are in agreement with the 
experiment  data with accuracy of 30-40 percents for baryonic masses and 
another static properties of baryons.

But there are many other ways for stabilization of chiral soliton, because
from the methodological point of view, arising of the Skyrme part in the
lagrangian of effective meson field theory is an ''ad hoc" procedure, and 
there 
are no real physical bases for this procedure. Of course, one can treat 
this part in the 
effective lagrangian  as a leading high derivative expansion of 
effective chiral meson action, but this approach 
leads to numerous questions so far. Two most important ones concern the 
physical nature of the scale parameter in this theory and the 
influence 
of another terms of expansion  on existence and stability of chiral 
solitons.

The consistent effective low-energy meson's theory construction is a very 
complicated task that is closely connected to the quark's confinement 
problem and to the problem of spontaneous breaking of chiral 
symmetry. It is desirable that such theory should be 
Lorentz and chiral invariant and should have a set of  finite
energy stable solutions. In the case of electro-magnetic fields such kind 
of an effective low-energy theory is the well-known Born-Infeld model
\cite{born}. In our work \cite{chbi} we study the direct analogue of the 
Born-Infeld 
action for  chiral fields. 
By construction our model has no singular solutions and looks 
very attractive as a possible effective action for mesons fields. 
This model has a set of stable topological solitons. These solutions can 
be treated as baryons states in our model.

In the  well-known paper \cite{born}, Born and Infeld  proposed a 
non-linear covariant action for electro-magnetic fields with very 
attractive features.  Firstly, in the framework of BI theory the problem 
of singular self-energy of electron can be solved. In this theory the 
electron is a stable finite energy solution of the BI field equation with 
electric charge.  Second, the BI action has a scale parameter $\beta$. 
Using expansion by this parameter one reduces the BI action to the usual 
Maxwell form in the low-energy limit.

 \begin{figure}[t]
 \leavevmode
 \epsfbox{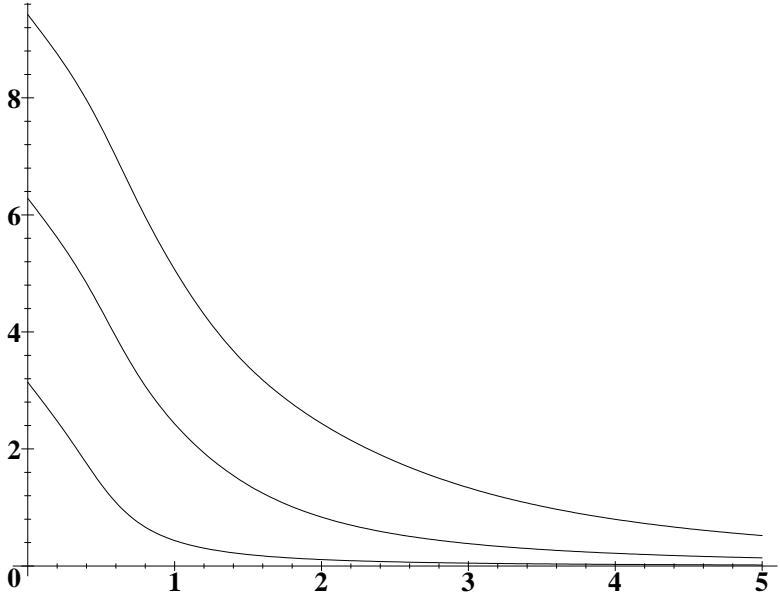}
 \caption{Solitons with $B=1, 2$ and $3$.
  Horizontal axis: r (in fm).}
 \end{figure}

We want  to perform a very similar procedure with the chiral prototype
lagrangian 
\beq
{\cal L}_{pr} = - \frac{f^2_\pi}{4} {\rm Tr} L_\mu L^\mu ,
\label{1.1}
\eeq
where $L_\mu=U^+ \ppd_\mu U$  is a Cartan left-invariant form and 
$ U=\exp (i \frac{\vec \phi_\pi \vec \tau}{f_\pi}) $ 
where the vector field $\vec \phi_\pi $ is associated with 
$\pi$ -  mesons and 
$f_\pi = 93 $ MeV is the pion decay constant.
Like in the case of BI action, our chiral model must have a set of  finite
energy solutions with integer values of charge (topological or baryon), and
in the low-energy limit such a theory must reproduce the prototype 
lagrangian (\ref{1.1}). The model must be 
Lorentz and chiral invariant. Finally, the
form of such theory follows from the analogy with the action of a 
relativistic particle, the BI action for EM and YM field. 

Arguing as above, let us consider a theory with lagrangian
\beq
{\cal L}_{ChBI} = - f^2_\pi {\rm Tr} 
\beta^2 \bigg(1-\sqrt{1-\frac 1{2\beta^2}L_\mu L^\mu } \bigg) 
\sim 
- \frac{f^2_\pi}{4} {\rm Tr} L_\mu L^\mu ,
\label{2.1}
\eeq
where $\beta$ is a mass dimensional scale parameter of our model. It is
easily shown that the expansion of the lagrangian (\ref{2.1}) gives us 
the prototype 
theory as the leading order theory by parameter $\beta$.

 \begin{figure}[t]
 \leavevmode
 \epsfbox{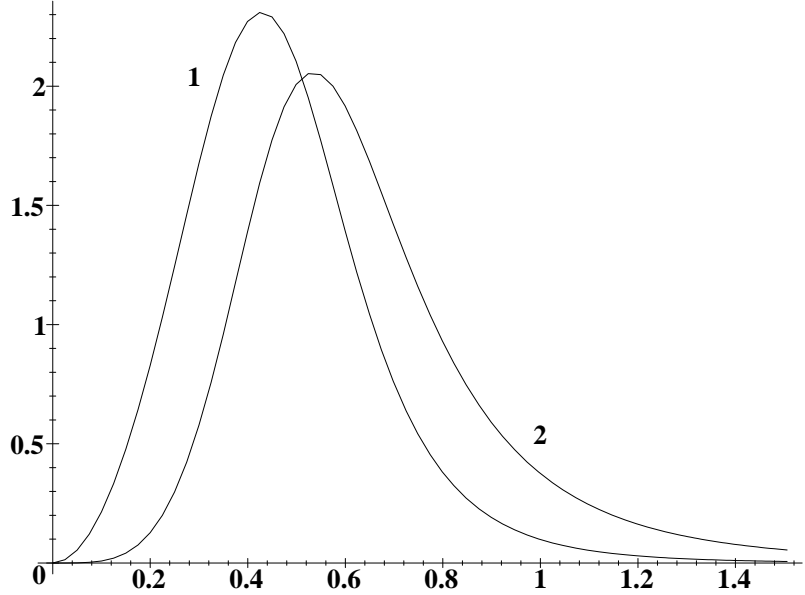}
 \caption{Baryonic density (1) and magnetic moment density (2).
  Horizontal axis: r (in fm).}
 \end{figure}

Now we consider the spherically symmetrical field
configuration
$ U=\exp ( i F(r) (\vec n \vec \tau))$, $ \vec n = \vec r/|r| $.
Using the variation principle, we get the
equation of motion
$$
(r^2 - \frac 1{\beta^2} \sin^2 F) F'' + ( 2rF' -\sin 2F) -
$$
\beq
- \frac 1{\beta^2} (r F'^3 - F'^2 \sin 2F + 3 \frac 1{r} F' \sin^2F 
- \frac 1{r^2} \sin2F \sin^2 F)=0.
\label{2.4}
\eeq

The numerical investigation of the solutions of equation (\ref{2.4}) 
is presented in Fig.1 (solitons with $B=1, 2$ and $3$).  
The scale parameter $\beta=807 \, \mbox{MeV}$ is defined from the 
hypothesis that the soliton  with 
$B=1$ is a nucleon. Indeed, using now a scale transformation, 
one gets 
$
\beta = 8 \pi f^2_\pi  E( B=1, \, \beta=1)/{m_p}=807 \, 
\mbox{MeV} $ , where $E( B=1, \, \beta=1)=3.487$ is the energy of this 
soliton solution for $B=1$ and $\beta=1$. 

Using this value of the $\beta$, one can find another physical properties
of the hadron. The baryonic density (see Fig.2) and the main radius of 
the hadron can 
be calculate by
$$
\rho_B = -\frac 2{pi} \sin^2 F \, F' \, , \, \, \, \, \, 
<r^2> = \int^\infty_0 r^2 \rho_B(r) dr .
$$  
For Chiral Born-Infeld model we have $<r^2>^{\mbox{\small ChBI}}\simeq 0.51 
\mbox{fm}$
(Skyrme model: $<r^2>^{\mbox{\small Sk}}\simeq 0.59 \mbox{\small fm}$); 
experiment data:  $<r^2>^{\mbox{\small exp}}\simeq 0.72 \mbox{fm}$ 
\cite{Wit} ).  The magnetic moment density (see Fig.2) and magnetic main 
radius of the hadron are $$ \rho_M = \frac{r^2 \sin^2 F \, 
F'}{\int^\infty_0 r^2 \sin^2 F \, F' dr} \, , \, \,\, \, \, <r^2>_M = 
\int^\infty_0 r^2 \rho_M(r) dr .  $$ For Chiral Born-Infeld model we have 
$<r^2>_M^{\mbox{\small ChBI}}\simeq 0.74 \mbox{fm}$ (Skyrme model: 
$<r^2>^{\mbox{\small Sk}}\simeq 0.92 \mbox{fm}$); experiment data:  
$<r^2>^{\mbox{\small exp}}\simeq 0.81 \mbox{fm}$ \cite{Wit}).

Using asymptotic of our solutions at infinity, one can find a prediction 
for the value of axial coupling constant $g_A^{\mbox{\small ChBI}}=0.79$. 
Such prediction lies much close to the experiment data $g_A^{\mbox{\small 
exp}}=1.23$ then the prediction of Skyrme model $g_A^{\mbox{\small 
Sk}}=0.61$ \cite{Wit}.   

We do not give a comprehensive investigation of  the Chiral
Born-Infeld theory. The questions about non-spherical solutions,
quantum fluctuations about such solutions and corresponding 
properties of baryons or about the nucleon-nucleon interaction  
are not clear now. But maybe the most
important question in such investigation is 
about physical substantiation of such theory. All of these questions 
should be the themes for a future investigation.

\end{document}